\documentclass[12pt,a4paper]{article}
\usepackage{amssymb,amsmath}
\usepackage[dvips]{lscape,graphicx}

\newcommand{\bc}{\begin{center}}
\newcommand{\ec}{\end{center}}
\newcommand{\bd}{\begin{displaymath}}
\newcommand{\ed}{\end{displaymath}}
\newcommand{\be}{\begin{equation}}
\newcommand{\ee}{\end{equation}}
\newcommand{\ba}{\begin{array}}
\newcommand{\ea}{\end{array}}
\newcommand{\bt}{\begin{tabular}}
\newcommand{\et}{\end{tabular}}

\newcommand{\ds}{\displaystyle}

\begin{document}

\title{Cosmological constant in SUGRA models with Planck scale
SUSY breaking and degenerate vacua}

\author{C.~D.~Froggatt${}^{a}$,
R.~Nevzorov${}^{b}$\footnote{On leave of absence from the Theory Department, SSC RF ITEP of NRC "Kurchatov Institute", Moscow, Russia.},
H.~B.~Nielsen${}^{c}$,
A.~W.~Thomas${}^{b}$\\[5mm]
\itshape{$^a$ School of Physics and Astronomy, University of Glasgow, Glasgow, UK}\\[0mm]
\itshape{$^b$ ARC Centre of Excellence for Particle Physics at the Terascale and CSSM,}\\[0mm]
\itshape{School of Chemistry and Physics, The University of Adelaide,}\\[0mm]
\itshape{$^c$ The Niels Bohr Institute, University of Copenhagen, Copenhagen, Denmark}}

\date{}

\maketitle

\begin{abstract}{
\noindent
The empirical mass of the Higgs boson suggests small to vanishing values of the quartic
Higgs self--coupling and the corresponding beta function at the Planck scale, leading to
degenerate vacua. This leads us to suggest that the measured value of the cosmological
constant can originate from supergravity (SUGRA) models with degenerate vacua.
This scenario is realised if there are at least three exactly degenerate vacua. In the first
vacuum, associated with the physical one, local supersymmetry (SUSY) is broken near the
Planck scale while the breakdown of the $SU(2)_W\times U(1)_Y$ symmetry takes
place at the electroweak (EW) scale. In the second vacuum local SUSY breaking
is induced by gaugino condensation at a scale which is just slightly lower than
$\Lambda_{QCD}$ in the physical vacuum. Finally, in the third vacuum local SUSY
and EW symmetry are broken near the Planck scale.}
\end{abstract}

\newpage
\section{Introduction}

The observation of the Higgs boson with a mass around $\sim 125-126$~GeV,
announced by the ATLAS \cite{:2012gk} and CMS \cite{:2012gu} collaborations
at CERN, is an important step towards our understanding of the mechanism of
the electroweak (EW) symmetry breaking. It is also expected that further exploration
of TeV scale physics at the LHC may lead to the discovery of new physics phenomena
beyond the Standard Model (SM) that can shed light on the stabilisation of the EW scale.
In the Minimal Supersymmetric (SUSY) Standard Model (MSSM) based on the softly
broken SUSY the scale hierarchy is stabilized because of the cancellation of quadratic
divergences (for a review see \cite{Chung:2003fi}).
The unification of gauge coupling constants, which takes place in SUSY models at high energies \cite{5},
allows the SM gauge group to be embedded into Grand Unified Theories (GUTs) \cite{4}
based on gauge groups such as $SU(5)$, $SO(10)$ or $E_6$. However, the cosmological
constant in SUSY extensions of the SM diverges quadratically and excessive fine-tuning
is required to keep its size around the observed value \cite{6}. Theories with flat \cite{7}
and warped \cite{8} extra spatial dimensions also allow one to explain the hierarchy
between the EW and Planck scales, providing new insights into gauge coupling
unification \cite{9} and the cosmological constant problem \cite{10}.

Despite the compelling arguments for physics beyond the SM, no signal or indication of
its presence has been detected at the LHC so far. Of critical importance here is the observation
that the mass of the Higgs boson discovered at the LHC is very close to the lower bound on
the Higgs mass in the SM that comes from the vacuum stability constraint \cite{1}-\cite{2}.
In particular, it has been shown that the extrapolation of the SM couplings up to the Planck
scale leads to (see \cite{Buttazzo:2013uya})
\begin{equation}
\lambda(M_{Pl})\simeq 0 \,, \qquad\quad \beta_{\lambda}(M_{Pl})\simeq 0\,,
\label{2}
\end{equation}
where $\lambda$ is the quartic Higgs self--coupling and $\beta_{\lambda}$ is its beta--function.
Eqs.~(\ref{2}) imply that the Higgs effective potential
has two rings of minima in the Mexican hat with the same vacuum energy density \cite{12}.
The radius of the little ring equals the EW vacuum expectation value (VEV) of the Higgs field,
whereas in the second vacuum $\langle H\rangle \sim M_{Pl}$.

The presence of such degenerate vacua was predicted \cite{12} by the so-called Multiple Point
Principle (MPP) \cite{mpp}-\cite{mpp-nonloc}, according to which Nature chooses values of coupling constants
such that many phases of the underlying theory should coexist. This scenario corresponds to
a special (multiple) point on the phase diagram of the theory where these phases meet.
The vacuum energy densities of these different phases are degenerate at the multiple point.
In previous papers the application of the MPP to the two Higgs doublet extension of the SM
was considered \cite{2hdm-1}--\cite{2hdm-2}. In particular, it was argued that the MPP
can be used as a mechanism for the suppression of the flavour changing neutral current and
CP--violation effects \cite{2hdm-2}.

The success of the MPP in predicting the Higgs mass \cite{12} suggests that we might also
use it for explaining the extremely low value of the cosmological constant. In particular,
the MPP has been adapted to models based on $(N=1)$ local supersymmetry -- supergravity
(SUGRA) \cite{Froggatt:2003jm}--\cite{Froggatt:2005nb}. As in the present article, we used
the MPP assuming the existence of a vacuum in which the low--energy limit of the theory is
described by a pure SUSY model in flat Minkowski space. Then the MPP implies that the physical
vacuum and this second vacuum have the same vacuum energy densities.
Since the vacuum energy density of supersymmetric states in flat Minkowski space
is just zero, the cosmological constant problem is thereby solved to first approximation.

However, the supersymmetry in the second vacuum can be broken dynamically when the
SUSY gauge interaction becomes non-perturbative at the scale $\Lambda_{SQCD}$, resulting
in an exponentially suppressed value of the cosmological constant which is then transferred to
the physical vacuum by the assumed degeneracy \cite{Froggatt:2003jm}--\cite{Froggatt:2005nb}.
A new feature of the present article is that we arrange for the hidden sector gauge interaction
to give rise to a gaugino condensate near the scale $\Lambda_{SQCD}$. This condensate then
induces SUSY breaking at an appreciably lower energy scale, via non-renormalisable terms.
The results of our analysis indicate that the appropriate value of the cosmological constant
in the second vacuum can be induced if  $\Lambda_{SQCD}$ is rather close to $\Lambda_{QCD}$,
that is near the scale where the QCD interaction becomes strong in the physical vacuum.

In this paper we also argue that both the tiny value of the dark energy density and
the small values of $\lambda(M_{Pl})$ and $\beta_{\lambda}(M_{Pl})$ can be incorporated
into the (N=1) SUGRA models with degenerate vacua. This requires that SUSY is not
broken too far below the Planck scale in the physical vacuum and that there exists a
third vacuum, which has the same energy density as the physical and second vacuum.
In this third vacuum local SUSY and EW symmetry should be broken near the Planck scale.

Our attempt to estimate the small deviation of the cosmological constant
from zero relies on the assumption that the physical and SUSY Minkowski
vacua are degenerate to very high accuracy. Although in the next section
we argue that in the framework of the $(N=1)$ supergravity the supersymmetric
and non--supersymmetric Minkowski vacua can be degenerate, it does not
shed light on the possible mechanism by which such an accurate
degeneracy may be maintained. In principle, a set of approximately
degenerate vacua can arise if the underlying theory allows only
vacua which have similar order of magnitude of space-time 4-volumes at the
final stage of the evolution of the Universe\footnote{This may imply the possibility
of violation of a principle that future can have no influence on the past \cite{mpp-nonloc}.}.
Since the sizes of these volumes are determined by the expansion rates of
the corresponding vacua associated with them, only vacua with similar order
of magnitude of dark energy densities are allowed. Thus all vacua are degenerate
to the accuracy of the value of the cosmological constant in the physical vacuum.

The paper is organized as follows: In the next section we specify an $(N=1)$ SUGRA scenario
that leads to the degenerate vacua mentioned above. In sections 3 and 4 we estimate the
dark energy density in such a scenario and discuss possible implications for Higgs
phenomenology. Our results are summarized in section 5.

\section{SUGRA models with degenerate vacua}

One may expect that at ultra--high energies the SM would be embedded in an underlying theory that
provides a framework for the unification of all interactions, including gravity, such as supergravity (SUGRA).
The full $(N=1)$ SUGRA Lagrangian \cite{04} is specified in terms of an analytic gauge
kinetic function $f_a(\phi_{M})$ and a real gauge-invariant K$\Ddot{a}$hler function
$G(\phi_{M},\phi_{M}^{*})$, which depend on the chiral superfields $\phi_M$. The function
$f_{a}(\phi_M)$ determines the gauge coupling constants $Re f_a(\phi_M)=1/g_a^2$, where the
index $a$ designates different gauge groups. The K$\Ddot{a}$hler function is a combination
of two functions
\be
G(\phi_{M},\phi_{M}^{*})=K(\phi_{M},\phi_{M}^{*})+\ln|W(\phi_M)|^2\,,
\label{3}
\ee
where $K(\phi_{M},\phi_{M}^{*})$ is the K$\Ddot{a}$hler potential whereas $W(\phi_M)$ is the
complete superpotential of the SUGRA model. Here we use standard supergravity mass units:
$\ds\frac{M_{Pl}}{\sqrt{8\pi}}=1$.

In order to obtain the vacuum, which is globally supersymmetric
with zero energy density and supersymmetry unbroken in first
approximation, we can just assume that the superpotential $W(\phi_M)$ and
its derivatives vanish near the corresponding minimum of the SUGRA
scalar potential \cite{Froggatt:2003jm}--\cite{Froggatt:2005nb}.
The simplest K$\Ddot{a}$hler potential and superpotential that satisfy
these conditions can be written as
\be
K(z,\,z^{*})=|z|^2\,,\qquad\qquad W(z)=m_0(z+\beta)^2\,.
\label{51}
\ee
The hidden sector of this SUGRA model contains only one singlet
superfield $z$. If the parameter $\beta=\beta_0=-\sqrt{3}+2\sqrt{2}$,
the corresponding SUGRA scalar potential possesses
two degenerate minima with zero energy density at the classical
level. One of them is a supersymmetric Minkowski minimum that
corresponds to $z^{(2)}=-\beta$. In the other minimum of the
SUGRA scalar potential ($z^{(1)}=\sqrt{3}-\sqrt{2}$)
local supersymmetry is broken; so it can be associated
with the physical vacuum. Varying the parameter $\beta$ around
$\beta_0$ one can obtain a positive or a negative contribution
from the hidden sector to the total energy density of the physical
vacuum. Thus $\beta$ can be fine--tuned so that the
physical and second vacua are degenerate.

In general, Eq.~(\ref{51})  represents the extra fine-tuning associated
with the presence of the supersymmetric Minkowski vacuum.
This fine-tuning can be to some extent alleviated in the no--scale inspired SUGRA models
with broken dilatation invariance \cite{Froggatt:2005nb}. Let us consider
a model with two hidden sector supermultiplets $T$ and $z$.
These superfields transform differently under the imaginary translations
($T\to T+i\beta,\, z\to z$) and dilatations ($T\to\alpha^2 T,\, z\to\alpha\,z$).
If the superpotential and K$\Ddot{a}$hler potential of the hidden sector
of the SUGRA model under consideration are given by
\begin{equation}
\begin{array}{rcl}
K(T,\,z)&=&\ds-3\ln\biggl[T+\overline{T}-|z|^2\biggr]\,,\\[2mm]
W(z)&=&\ds\kappa\biggl(z^3+ \mu_0 z^2 \biggr)\,,
\end{array}
\label{61}
\end{equation}
then the corresponding tree level scalar potential of the hidden sector
is positive definite
\begin{equation}
V(T,\, z)=\frac{1}{3(T+\overline{T}-|z|^2)^2}
\biggl|\frac{\partial W(z)}{\partial z}\biggr|^2\,,
\label{6}
\end{equation}
so that the vacuum energy density vanishes near its global minima.
The scalar potential (\ref{6}) possesses two minima  at $z=0$ and
$z=\ds-\frac{2\mu_0}{3}$ that correspond to the stationary points of
the hidden sector superpotential. In the first vacuum, where
$z=\ds-\frac{2\mu_0}{3}$, local supersymmetry is broken so that
the gravitino becomes massive
\begin{equation}
m_{3/2}=\biggl<\frac{W(z)}{(T+\overline{T}-|z|^2)^{3/2}}\biggr>
=\frac{4\kappa\mu_0^3}{27\biggl<\biggl(T+\overline{T}
-\frac{4\mu_0^2}{9}\biggr)^{3/2}\biggr>}\,.
\label{62}
\end{equation}
and all scalar particles get non--zero masses. Since one can
expect that $\mu_0\lesssim M_{Pl}$ and $\kappa\lesssim 1$
SUSY is broken in this vacuum near the Planck scale. In the second
minimum, with $z=0$, the superpotential of the hidden sector vanishes
and local SUSY remains intact, so that the low--energy limit of
this theory is described by a pure SUSY model in flat Minkowski
space.

Of course, the inclusion of perturbative and non--perturbative corrections to the
Lagrangian of the no--scale inspired SUGRA model, which should depend on the
structure of the underlying theory, are expected to spoil the degeneracy of vacua
inducing a huge energy density in the vacuum where SUSY is broken. Moreover
in this SUGRA model the mechanism for the stabilization of the vacuum expectation
value of the hidden sector field $T$ remains unclear. The model discussed above
should therefore be considered as a toy example only. This SUGRA model demonstrates that,
in $(N=1)$ supergravity, there might be a mechanism which ensures the vanishing
of vacuum energy density in the physical vacuum. This mechanism may also lead to
a set of degenerate vacua with broken and unbroken supersymmetry, resulting in
the realization of the multiple point principle.

\section{Cosmological constant and $\Lambda_{QCD}$}

We now assume that a phenomenologically viable SUGRA model with degenerate vacua
of the type just discussed is realised in Nature. In other words we assume that there are at
least two vacua which are exactly degenerate. In the first (physical) vacuum SUSY is broken
near the Planck scale. In the second vacuum SUSY remains intact. We shall assume that,
by one way or another, only vector supermultiplets, which correspond to the unbroken gauge
symmetry in the hidden sector, remain massless. These supermultiplets, that survive
to low energies, give rise to the breakdown of SUSY
in the second vacuum
which is caused by the formation of a
gaugino condensate induced in the hidden sector at the scale $\Lambda_{SQCD}$ much
lower than $M_{Pl}$. However, since the gaugino condensate does not actually break global
SUSY \cite{027}, it is only via the effect of a non-renormalisable term that this condensate
causes the SUSY breaking. As a consequence the SUSY breaking scale is many orders of
magnitude lower than $\Lambda_{SQCD}$. Here we assume that other fields, such as
the visible SM fields, for example, give a much smaller contribution to the energy density of the second
vacuum  than the hidden pure super Yang Mills fields\footnote{This can be achieved, for example,
if the gauge kinetic function associated with the ordinary QCD interactions in the visible sector
is sufficiently large in the second vacuum.}.

In order to give a formulation of the non-renormalisable effect that eventually
facilitates the breakdown of SUSY, we remark that we can have a non-trivial dependence
of the gauge kinetic function $f_{X}(h_m)$ on the hidden sector superfields $h_m$.
Such a dependence leads to auxiliary fields corresponding to the hidden fields $h_m$
\be
F^{h_m}\propto \frac{\partial f_X(h_k)}{\partial h_m}\bar{\lambda}_a\lambda_a+...
\label{16}
\ee
acquiring non--zero VEVs, which are set by $<\bar{\lambda}_a\lambda_a>\simeq \Lambda_{SQCD}^3$.
This results in supersymmetry breaking \cite{020} at the scale $M_S^2 \sim \frac{\Lambda^3_{SQCD}}{M_{Pl}}$
and a non--zero vacuum energy density
\be
\rho^{(2)}_{\Lambda} \sim M_S^4 \sim \frac{\Lambda_{SQCD}^6}{M_{Pl}^2}\, .
\label{17}
\ee

\begin{figure}
\begin{center}
{\includegraphics[height=115mm,keepaspectratio=true]{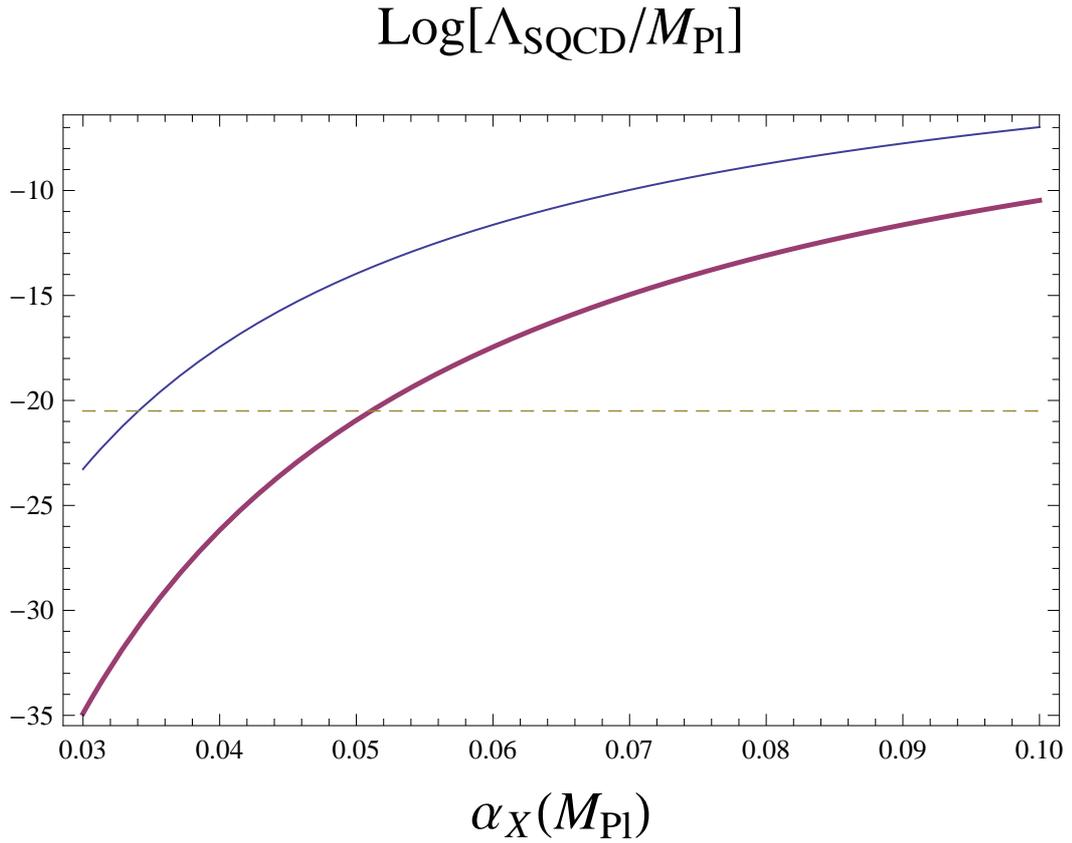}}
\end{center}
\caption{The value of $\log\left[\Lambda_{SQCD}/M_{Pl}\right]$ versus $\alpha_X(M_{Pl})$.
The thin and thick solid lines correspond to the $SU(3)$ and $SU(2)$ gauge symmetries, respectively.
The horizontal line is associated with the value of $\Lambda_{SQCD}$ that leads to the observed
value of the cosmological constant.}
\label{new-fig1}
\end{figure}

The postulated exact degeneracy of vacua implies then that the physical vacuum, in which
SUSY is broken near the Planck scale, has the same energy density as the phase where local
supersymmetry breakdown takes place at $\Lambda_{SQCD}$. Using Eq.~(\ref{17}) one
easily finds that, in order to reproduce the observed value of the cosmological constant,
$\Lambda_{SQCD}$ should be relatively close to $\Lambda_{QCD}$ in the physical vacuum,
i.e.
\be
\Lambda_{SQCD}\sim \Lambda_{QCD}/10\,.
\label{18}
\ee
Although there is no compelling theoretical reason to expect a priori that the two scales
$\Lambda_{SQCD}$ and $\Lambda_{QCD}$ should be relatively close or related,
one might naively consider $\Lambda_{QCD}$ and $M_{Pl}$ as the two most natural choices
for the scale of dimensional transmutation in the hidden sector.

Using the analytical solution of the one--loop RG equation
\be
\ds\frac{1}{\alpha_X(Q)}=\ds\frac{1}{\alpha_X(M_{Pl})}+\frac{b_X}{4\pi}\ln\frac{M^2_{Pl}}{Q^2}\,,
\label{19}
\ee
one can estimate the energy scale, $\Lambda_{SQCD}$, where the supersymmetric QCD-like interactions
become strong in the second vacuum, for a given value of $\alpha_X(M_{Pl})$. In Eq.~(\ref{19})
$b_X=-9$ and $-6$ for the $SU(3)$ and $SU(2)$ gauge groups respectively.
Setting
$\ds\frac{1}{\alpha_X(\Lambda_{SQCD})}\to 0$ one finds
\be
\Lambda_{SQCD}=M_{Pl}\exp\left[{\frac{2\pi}{b_X \alpha_X(M_{Pl})}}\right]\,.
\label{20}
\ee
The dependence of $\Lambda_{SQCD}$ on $\alpha_X(M_{Pl})$ is shown in
Fig.~1. As one expects, the value of $\Lambda_{SQCD}$ diminishes with decreasing
$\alpha_X(M_{Pl})$. The measured value of the cosmological constant is reproduced when
$\alpha_X(M_{Pl})\simeq 0.051$ in the case of the model based on the $SU(2)$ gauge group and
$\alpha_X(M_{Pl})\simeq 0.034$ in the case of the $SU(3)$ SUSY gluodynamics. These values
of $\alpha_X(M_{Pl})$ correspond to $g_X(M_{Pl})\simeq 0.801$ and $g_X(M_{Pl})\simeq 0.654$
respectively. Thus in the case of the $SU(3)$ model the gauge coupling $g_X(M_{Pl})$
is just slightly larger than the value of the QCD gauge coupling at the Planck scale, i.e.
$g_3(M_{Pl})=0.487$ (see Ref.~\cite{Buttazzo:2013uya}), in the physical vacuum where we live.

\section{Preserving the Higgs mass prediction}

Now we shall consider the implications of SUGRA models with degenerate vacua
for Higgs phenomenology. The presence of two vacua, as discussed above, does not rule out the
possibility that there might be other vacua with the same energy density too. In particular, there
can exist a vacuum where EW symmetry is broken near the Planck scale. Because the
Higgs VEV is somewhat close to $M_{Pl}$ one must consider the interaction of the Higgs
and hidden sector fields. Thus the full scalar potential can be written:
\begin{equation}
V=V_{hid}(h_m) + V_0(H) + V_{int}(H, h_m)+...\,,
\label{21}
\end{equation}
where $V_{hid}(h_m)$ is the part of the scalar potential associated with the hidden sector,
$V_0(H)$ is the part of the full scalar potential that depends on the Higgs field only and
$V_{int}(H, h_m)$ corresponds to the interaction of the SM Higgs field with the hidden sector fields.
In what follows we assume that in the observable sector only one Higgs doublet acquires
a non--zero VEV, so that all other observable fields can be ignored in the first approximation.

In general one expects that in the vacuum with the Planck scale VEV of the Higgs doublet
the VEVs of the hidden sector fields should be very different from those in the physical
vacuum. As a consequence, in this third vacuum the gauge couplings at the Planck
scale, as well $\lambda(M_{Pl})$ and $m^2(M_{Pl})$, are not the same as in the physical
vacuum. Moreover, it seems to be basically impossible to establish any model independent
relation between the values of these couplings in different vacua in general. However, in the limit when
$V_{int}(H, h_m)\to 0$ the situation changes. In this case, the Planck scale VEV of the Higgs
field would not lead to substantial variations of the VEVs of hidden sector fields.
Thus the gauge couplings and $\lambda(M_{Pl})$ in the third and physical vacua would be almost
identical. Then the requirement of the degeneracy of all three vacua leads to the conditions (\ref{2}).

Although it is difficult to justify why $V_{int}(H, h_m)$ should be vanishingly small,
the interactions between the SM Higgs doublet and the hidden sector fields can be rather weak
near the third vacuum, i.e. $V_{int}(H, h_m)\ll M_{Pl}^4$. This may happen, for example, if
the VEV of the Higgs field is considerably smaller than $M_{Pl}$ (say $<H>\sim M_{Pl}/10$)
and the couplings of the SM Higgs doublet to the hidden sector fields are suppressed.
In this case the relatively large Higgs VEV associated with the third vacuum may not
affect much the VEVs of the hidden sector fields, so that the gauge couplings and $\lambda(M_{Pl})$
remain almost the same as in the physical vacuum. At the same time, the absolute value of
$m^2$ in the Higgs effective potential should  - although fixed to be small at the weak scale according
to   experiment - be much larger in the third vacuum. Indeed, in the
physical vacuum this parameter might well be small because of the cancellation of different
contributions. However, even small variations of the VEVs of the hidden sector fields are
expected to spoil such cancellations in general. Nonetheless, if the interactions between the
SM Higgs doublet and hidden sector fields are weak, $m^2(M_{Pl})$ can still be substantially
smaller than $M_{Pl}^2$ and $\langle H^{\dagger} H\rangle$ in the third vacuum. If indeed
$V_{int}(H, h_m)\ll M_{Pl}^4$ and the VEVs of the hidden sector fields do not change much,
i.e. the value of $V_{hid}(h_m)$ also remains almost the same as in the physical vacuum
where $V_{hid}(h^{(1)}_m)\ll M_{Pl}^4$, the requirement of the degeneracy of vacua implies
that in the third vacuum $\lambda(M_{Pl})$ and $\beta_{\lambda}(M_{Pl})$ are approximately zero.
Because in this case the couplings in the third and physical vacua are basically identical,
the presence of such a third vacuum results in the predictions (\ref{2}) for $\lambda(M_{Pl})$ and
$\beta_{\lambda}(M_{Pl})$ in the physical vacuum.

Since we do not have EW scale SUSY and have
already assumed MPP, we might imagine incorporating yet a fourth vacuum \cite{hierarchy, sixtop}
into our model, which could then provide an MPP fine-tuning solution to the hierarchy problem.
Although this is an interesting possibility the corresponding analysis goes well beyond the
scope of this paper.

\section{Conclusions}

In this note, inspired by the observation that the mass of the recently discovered Higgs boson leads
naturally to Eq.~(\ref{2}) and degenerate vacua in the Standard Model, we have argued that SUGRA models
with degenerate vacua can lead to a rather small dark energy density, as well as small values of
$\lambda(M_{Pl})$ and $\beta_{\lambda}(M_{Pl})$.
This is realised in a scenario where the existence of at least three exactly degenerate vacua is postulated.
In the first (physical) vacuum SUSY is broken near the Planck scale and the small value of the
cosmological constant appears as a result of the fine-tuned precise cancellation of different
contributions. In the second vacuum the breakdown of local supersymmetry is induced by gaugino
condensation, which is formed at the scale $\Lambda_{SQCD}$ where hidden sector gauge interactions
become strong. If $\Lambda_{SQCD}$ is slightly lower than $\Lambda_{QCD}$ in the physical vacuum,
then the energy density in the second vacuum is rather close to $10^{-120}M_{Pl}^4$. Because of the
postulated degeneracy of vacua, this tiny value of the energy density is transferred to the other vacua
including the one where we live. In the case of the hidden sector gauge group being $SU(3)$, the
measured value of the cosmological constant \cite{6} is reproduced for
a value of $\alpha_X(M_{Pl})$ which is only
slightly above that of the strong gauge coupling at the Planck scale in the physical vacuum.

Finally, the presence of the third degenerate vacuum, where local SUSY and EW symmetry are broken
somewhere near the Planck scale, can constrain $\lambda(M_{Pl})$ and $\beta_{\lambda}(M_{Pl})$
in the physical vacuum. This may happen if the VEV of the Higgs field is considerably smaller than
$M_{Pl}$ (say $\langle H\rangle \lesssim M_{Pl}/10$). Then the large Higgs VEV may not affect much
the VEVs of the hidden sector fields. As a consequence $m^2$ in the Higgs effective potential is
expected to be much smaller than $M_{Pl}^2$ and $\langle H^{\dagger} H\rangle$ in the third vacuum.
Thus the existence of such a third vacuum with vanishingly small energy density would still imply that
$\lambda(M_{Pl})$ and $\beta_{\lambda}(M_{Pl})$ are approximately zero in this vacuum.
Since we are taking the VEVs of the hidden sector fields to be almost identical in the physical and
third vacua, we also expect $\lambda(M_{Pl})$ and $\beta_{\lambda}(M_{Pl})$ to be almost the same.
Consequently we obtain $\lambda(M_{Pl})\approx \beta_{\lambda}(M_{Pl})\approx 0$ in the physical vacuum.

It is worth noting that our estimate of the tiny value of the cosmological constant
makes sense only if the vacua mentioned above are degenerate to very high accuracy.
The identification of a mechanism that can give rise to a set
of vacua which are degenerate to such high accuracy is still a work in progress.
Here we just remark that vacua with very different dark energy densities
should result in very different expansion rates and ultimately in very different space--time
volumes for the Universe. If the underlying theory allows only vacua which lead to the
similar order of magnitude of space-time 4-volumes then such vacua should be
degenerate to the accuracy of the value of the dark energy density in the physical
vacuum.

\vspace{-5mm}
\section*{Acknowledgements}
\vspace{-3mm}
R.N. is grateful to J.~Bjorken, D.~Gorbunov, M.~Libanov, V.~Novikov, O.~Kancheli, D.~Kazakov, S.~F.~King,
S.~Pakvasa, V.~Rubakov, M.~Sher, D.~G.~Sutherland, S.~Troitsky, X.~Tata, M.~Vysotsky for  fruitful discussions.
This work was supported by the University of Adelaide and the Australian Research Council through the ARC
Center of Excellence in Particle Physics at the Terascale and through grant LFO 99 2247 (AWT). HBN thanks
the Niels Bohr Institute for his emeritus status. CDF thanks Glasgow University and the Niels Bohr Institute
for hospitality and support.

\end{document}